\documentclass[twocolumn,aps,pra,superscriptaddress,nofootinbib]{revtex4-2}

\usepackage{pifont}
\newcommand{\cmark}{\ding{51}}%
\newcommand{\xmark}{\ding{55}}%

\usepackage{graphicx}
\usepackage[dvipsnames,svgnames]{xcolor}
\usepackage{bm,bbm}
\usepackage{braket}
\usepackage{amsthm,amsmath,amssymb}
\usepackage{enumerate}
\usepackage{booktabs,multirow}
\usepackage{mathtools,bbding}
\usepackage[percent]{overpic}
\usepackage{microtype}
\usepackage{array,adjustbox,afterpage}
\usepackage[T1]{fontenc}
\usepackage[utf8]{inputenc}
\usepackage{tikz}
\usetikzlibrary{calc}
\usepackage[english]{babel}
\usepackage{newtxtext,newtxmath}

\usepackage{lipsum}

\usepackage[linesnumbered,ruled]{algorithm2e}  

\makeatletter
\renewcommand{\fnum@figure}{FIG. \thefigure} 
\makeatother

\usepackage[bookmarks=false, colorlinks=true, linkcolor=blue, citecolor=purple, urlcolor=purple]{hyperref}
\usepackage{cleveref}
\usepackage[normalem]{ulem}

\usepackage{lipsum}

\Crefname{subfigures}{figure}{figures}
\Crefname{subfigures}{Figure}{Figures}

\def\bra#1{\mathinner{\langle{#1}|}}
\def\ket#1{\mathinner{|{#1}\rangle}}

\def\ketbra#1#2{|{#1}\rangle\langle{#2}|}

\newcommand{\xdownarrow}[1]{{\left\downarrow\vbox to #1{}\right.\kern-\nulldelimiterspace}} 

\hypersetup{
    bookmarksnumbered=false, 
    breaklinks=true,
    unicode=false, 
    pdfstartview={FitH}, 
    pdftitle={}, 
    pdfnewwindow=true, 
}

\begin{document}

\title{What is \textit{Quantum} in Probabilistic Explanations of \\the Sure Thing Principle Violation?}

\author{Nematollah Farhadi Mahalli}
\email{nmahalli20@ku.edu.tr}
\author{Onur Pusuluk}
\thanks{corresponding author}
\email{onur.pusuluk@gmail.com}
\affiliation{Department of Physics, Ko{\c{c}} University, 34450 Sar{\i}yer, Istanbul, Turkey}%

\date{\today}

\begin{abstract}
The Prisoner's Dilemma game (PDG) is one of the simple test-beds for the probabilistic nature of the human decision-making process. Behavioral experiments have been conducted on this game for decades and show a violation of the so-called \textit{sure thing principle}, a key principle in the rational theory of decision. Quantum probabilistic models can explain this violation as a second-order interference effect, which cannot be accounted for by classical probability theory. Here, we adopt the framework of generalized probabilistic theories and approach this explanation from the viewpoint of quantum information theory to identify the source of the interference. In particular, we reformulate one of the existing quantum probabilistic models using density matrix formalism and consider different amounts of classical and quantum uncertainties for one player's prediction about another player's action in PDG. This enables us to demonstrate that what makes possible the explanation of the violation is the presence of \textit{quantum coherence} in the player's initial prediction and its conversion to probabilities during the dynamics. Moreover, we discuss the role of other quantum information-theoretical quantities, such as quantum entanglement, in the decision-making process. Finally, we propose a three-choice extension of the PDG to compare the predictive powers of quantum probability theory and a more general probabilistic theory that includes it as a particular case and exhibits third-order interference.

\begin{description}
\item[Keywords]
prisoner’s dilemma, sure thing principle, quantum coherence, generalized probabilistic theories
\end{description}
\end{abstract}

\maketitle

\section{Introduction}

Can we interpret the measurement statistics of a physical system for which we do not have sufficient knowledge of its nature and laws? Generalized probabilistic theories (GPTs)~\cite{pop94, Hardy01, Barrett07, Paterek10, Muller21} provide a mathematical framework for this aim and model a given experiment by decomposing it into labeled preparations, transformations, and measurements. For example, preparations are represented by probability vectors, whereas transformations and measurements are both described by stochastic matrices in classical probability theory (CPT). On the other hand, in quantum probability theory (QPT), density matrices stand for the preparations, their transformations are governed by completely-positive trace-preserving maps, and the operations called positive operator-valued measures are responsible for measurements.

Building on such abstract mathematical structures, the framework of GPTs enables the interrogation of the physical principles behind a given measurement statistic. Popescu and Rohrlich proposed for the first time in Ref.~\cite{pop94} that there should be a set of physically motivated axioms from which one can derive QPT. They focused on nonlocality and relativistic causality but concluded that these axioms produce more of what QPT can theorize~\cite{pop94, pop14}. To this aim, they characterized the extent of the violation of CHSH-type Bell inequality~\cite{Bell64, Clau69}, which is respectively constrained by $2$ and $2\sqrt{2}$ under CPT and QPT~\cite{Csir80, Csir87, Csir92}. However, it turned out that nonlocality and relativistic causality give an upper bound of $4$~\cite{pop94}, which means that they are necessary but insufficient for QPT. This finding also signifies that nonlocality and relativistic causality may lead to non-quantum or beyond-quantum GPTs with some notable nonclassical features of the QPT. A variety of choices for the axioms allows us to single out QPT from this entire set of non-quantum and beyond-quantum GPTs~\cite{Hardy01, Mas11, Rau11, Chiribella11}. Moreover, a weaker version of a set of axioms that singles outs QPT can also be compatible with CPT~\cite{Mas11}.

Besides this, the experiments in which the CPT fails to calculate the measurement statistic are not limited to the Bell inequality experiments. For instance, empirical data suggest that people appear to violate one of the basic tenets of rational decision theory, the so-called sure-thing principle~\cite{Savage54, Pearl16} while making a decision under uncertainty~\cite{Kah83, Tver92, Shafir92, Croson1999, Li2002}, e.g., during the Prisoner’s Dilemma game~\cite{Shafir92, Croson1999, Li2002}.  QPT models of the decision-making process can explain this violation as a quantum interference effect~\cite{Buse06, Buse09, Khrenn09, Tes21, Poth21}, which has also been used in decision theory to explain many other phenomena~\cite{Wang13, Aerts13, Buse16, Buse17, Khrren09, Basi21}. But what is the source of this interference, and how does it relate to the uncertainty? Here, we aim to develop a convenient approach to explore these questions by utilizing the framework of GPTs.

The paper is organized as follows: In Section~\ref{Sec::CPTQPT}, we will explain the appropriate representation of a generic mental state to model the violation of the sure thing principle using QPT. Section~\ref{Sec::AnalResuls} will provide an analytical demonstration that attributes the violation solely to the transformations between probabilities and quantum coherence in the initial prediction. The conceptual relationships among quantum uncertainty, quantum coherence, and quantum entanglement will be discussed in Section~\ref{Sec::QInfo}. Section~\ref{Sec::NumResuls} will present numerical results for various initial mental states. In Section~\ref{Sec::Outlook}, we will explore how the GPT framework can enhance existing QPT models and propose methods to validate their effectiveness.

\newpage

\section{Model System} \label{Sec::CPTQPT}

Without any loss of generality, we can model a decision-making process by focusing on the well-known two-player game \textit{Prisoner's dilemma}. In this game, a player’s decision does not depend solely on their rational reasoning. It is also related to their prediction about the other player's choice.

Assume that each player in the game has to pick one out of two actions denoted by \(c\) and \(d\). Then a player's decision-making process can be modelled using CPT as follows. The probability vector
\begin{equation}\label{Eq::JointP}
\vec{p} = \{p_{dd}, p_{dc}, p_{cd}, p_{cc}\} ,
\end{equation}
represents the mental state of the player. Here, \(p_{ij}\) stands for the joint probability \(p(B=i, A=j)\) with \(A\) and \(B\) respectively denoting the choice of the player and her/his prediction about the other player's choice. Marginal probabilities are related to joint probabilities through Bayes' theorem, which states that
\begin{eqnarray}\label{Eq::Bayes}
p(B=i, A = j) = p(A=j|B=i) \, p(B=i) \, ,
\end{eqnarray}
where \(p(A=j|B=i)\) is the conditional probability, the probability for the player to choose action \(j\) when she/he is sure the other player has chosen action \(i\). So, the action (prediction) state of the player can be obtained from the mental state~(\ref{Eq::JointP}) after tracing out the marginal prediction (action) probabilities by:
\begin{eqnarray}
\vec{p}_A &=& \{\sum_i p_{id}, \sum_i p_{ic}\} , \label{Eq::MarginalPA} \\
\vec{p}_B &=& \{\sum_j p_{dj}, \sum_j p_{cj}\} , \label{Eq::MarginalPB}
\end{eqnarray}
such that the first component of \(\vec{p}_A\) (\(\vec{p}_B\)) is equal to the marginal probability \(p(A=d)\) (\(p(B=d)\)).

The density matrix formalism allows us to consider CPT as a limiting case of QPT and recover all predictions made by the probability vectors. For example, the mental state given in Eq.~(\ref{Eq::JointP}) can be written as a \(4 \times 4\) diagonal density matrix
\begin{equation} \label{Eq::Rho::CC}
\rho = diag(\vec{p}) ,
\end{equation}
and its partial traces over \(B\) and \(A\) give the following action and prediction states
\begin{eqnarray}
\rho_A &\equiv& \mathrm{tr}_B[\rho] = diag(\vec{p}_A) , \label{Eq::RhoA::C} \\
\rho_B &\equiv& \mathrm{tr}_A[\rho] = diag(\vec{p}_B) . \label{Eq::RhoB::C}
\end{eqnarray}

On the other hand, QPT predicts more general states, such as
\begin{equation} \label{Eq::RhoA::Q}
\rho_A =
\begin{pmatrix}
p_A & \lambda_A \\
\lambda_A^* & 1 - p_A
\end{pmatrix} ,
\end{equation}
which is also a valid density matrix for the action state. Here, the diagonal elements correspond to the probabilities \(p_A = p(A=d)\) and \(1 - p_A = p(A=c)\). When \(p_A = |\psi_d|^2\), \(1 - p_A = |\psi_c|^2\), and \(\lambda_A = \psi_d \, \psi_c^*\), the state is called pure. That is to say, it can be written as the self-outer product of a state vector \(\ket{\psi}_A = \psi_d \ket{d}_A + \psi_c \ket{c}_A\), i.e., \(\rho_A = \ketbra{\psi}{\psi}_A\) with \(\bra{\psi}\) is the Hermitian conjugate of \(\ket{\psi}\) and
\begin{equation}
    \ket{d} =
    \begin{pmatrix}\label{Eq::dc basis}
    1 \\
    0
    \end{pmatrix} , \ \
    \ket{c} =
    \begin{pmatrix}
    0 \\
    1
    \end{pmatrix} .
\end{equation}

If \(\rho_A\) cannot be written as a self-outer product, it is called mixed, which means that the action state is a probabilistic mixture of two or more state vectors. The classical action state~(\ref{Eq::RhoA::C}) is an example of mixed states, e.g., it equals to \(p_A \ketbra{d}{d}_A + (1 - p_A) \ketbra{c}{c}_A\). However, \(|\lambda_A|^2 \leq \sqrt{p_A \, (1-p_A)}\) in general.

Similarly, a generic prediction state reads
\begin{equation} \label{Eq::RhoB::Q}
\rho_B =
\begin{pmatrix}
p_B & \lambda_B \\
\lambda_B^* & 1 - p_B
\end{pmatrix} ,
\end{equation}
where \(p_B = p(B=d)\), \(1 - p_B = p(B=c)\), and \(|\lambda_B|^2 \leq \sqrt{p_B \, (1-p_B)}\).

The presence of non-zero off-diagonal elements in Eqs.~(\ref{Eq::RhoA::Q})~and~(\ref{Eq::RhoB::Q}) is known as \textit{quantum coherence}. Although these elements have a non-probabilistic character, their interconversion with the diagonal elements is not prohibited in a general QPT transformation. So they can, in principle, contribute to the final probabilities in a decision-making process.

Whenever the total mental state of the player cannot be written in the form of
\begin{equation} \label{Eq::Rho::Ent}
\rho = \sum_k q_k \, \rho_B^{(k)} \otimes \rho_A^{(k)} ,
\end{equation}
it is called \textit{inseparable} or \textit{quantum entangled}. In other words, the action and prediction of the player share non-classical correlations known as \textit{quantum entanglement}. This quantum information-theoretical quantity is a special case of quantum coherence, which is not localized in subsystems but shared between them. We will revisit the concepts of quantum coherence and entanglement in more detail in Sec.~\ref{Sec::QInfo}.

\section{Analytical Results} \label{Sec::AnalResuls}

The probability vector given in Eq.~(\ref{Eq::JointP}) cannot describe the violation of the sure-thing principle (STP). When the player is sure that the other player has chosen \(i\), the probability for the player to choose action \(d\) becomes \(p_{A|B=i} \equiv p(A=d|B=i)\). Conversely, if there is uncertainty about the action of the other player, this probability turns out to be average, as below
\begin{eqnarray} \label{Eq::STPv1}
\begin{aligned}
p_{A|B=u} = p_B \, p_{A|B=d} + (1-p_B) \, p_{A|B=c} ,
\end{aligned}
\end{eqnarray}
where the subscript \(u\) stands for uncertain. This equation implies the STP: If \(p_{A|B=d} = p_{A|B=c} = 1\), then the player should choose to action \(d\) with a unit probability under uncertainty. However, according to the empirical observations~\cite{Shafir92, Croson1999, Li2002}, Eq.~(\ref{Eq::STPv1}) does not hold as \(p_{A|B=u}\) is generally less than any average of \(p_{A|B=d}\) and \(p_{A|B=c}\). Hence, people can violate the STP, at least in the experiments, which in turn makes the Bayesian CPT insufficient to describe a decision-making process.

Let us demonstrate how the STP can be violated in QPT. Assume that the mental state of the player at time \(t\) is represented by \(\rho^{(\alpha)}(t) = \varepsilon_t [\rho^{(\alpha)}(0)]\) with \(\varepsilon_t\) is a completely-positive trace-preserving map and \(\alpha=\{d,c,u\}\). Here, \(\alpha = u\) corresponds to a lack of exact prediction at the beginning of the decision-making process. Otherwise, the player has complete knowledge about the action the other player has chosen, that is, the initial prediction state is \(\mathrm{tr}_A[\rho^{(\alpha \neq u)}] = \ketbra{\alpha}{\alpha}_B\). Also, assume that the initial action state is the same for all values of \(u\).

It is straightforward to show the following relationship between the three different initial states
\begin{equation} \label{Eq::RhoU0}
    \rho^{(u)}(0) = p_B \, \rho^{(d)}(0) + (1 - p_B) \, \rho^{(c)}(0) + \chi(0) \, ,
\end{equation}
as long as the uncertainty does not require any classical or quantum correlations initially between the action and prediction, that is, \(\rho^{(u)}(0) = \rho_B(0) \otimes \rho_A(0)\). The last term above reads
\begin{equation} \label{Eq::Ksi0}
\chi(0) =
\begin{pmatrix}
0 & 0 & p_A \, \lambda_B & \lambda_A \, \lambda_B \\
0 & 0 & \lambda_A^* \, \lambda_B & (1 - p_A) \, \lambda_B \\
p_A \, \lambda_B^* & \lambda_A \, \lambda_B^* & 0 & 0 \\
\lambda_A^* \, \lambda_B^* & (1 - p_A) \, \lambda_B^* & 0 & 0
\end{pmatrix} ,
\end{equation}
and the linearity of \(\varepsilon_t\) preserves the form of the decomposition of \(\rho^{(u)}\) for all times
\begin{equation} \label{Eq::RhoU}
    \rho^{(u)}(t) = p_B \, \rho^{(d)}(t) + (1 - p_B) \, \rho^{(c)}(t) + \chi(t) \, .
\end{equation}

The trace of \(\chi(t)\) should be always zero as \(\mathrm{tr}[\rho^{(u)}(t)] = \mathrm{tr}[\rho^{(d)}(t)] = \mathrm{tr}[\rho^{(c)}(t)] = 1\). If all the elements in its diagonal are also zero, \(\chi(t)\) never contributes to the probabilities but only carries quantum coherence. This ensures that the probabilities \(p_{A|B=u}\), \(p_{A|B=d}\), and \(p_{A|B=c}\) obey the STP. On the other hand, non-vanishing diagonals of \(\chi(t)\) may violate the principle. Eq.~(\ref{Eq::RhoU}) can be used to identify the explicit relationship between the probabilities
\begin{equation}\label{Eq::STPv2}
    \begin{aligned}
    p_{A|B=u} = \, p_B \, p_{A|B=d} + (1 - p_B) \, p_{A|B=c} + \delta \, ,
    \end{aligned}
\end{equation}
with
\begin{equation}
     p_{A|B=\alpha} \equiv p_{A|B=\alpha}(t) = \sum_{i=d,c} \langle i_B \otimes d_A | \rho^{(\alpha)}(t) | i_B \otimes d_A \rangle \, ,
\end{equation}
and
\begin{equation}\label{Eq:Sure thing Viol}
\delta \equiv \delta(t) = \sum_{i=d,c} \langle i_B \otimes d_A | \chi(t) | i_B \otimes d_A \rangle \, .
\end{equation}

Eq.~(\ref{Eq::STPv2}) differs from (\ref{Eq::STPv1}) by the presence of \(\delta(t)\). Whenever this term has a non-zero value, the STP is violated. Hence, the necessary and sufficient condition to satisfy the STP during all dynamics is the absence of quantum coherence in the initial prediction state, i.e., \(\delta(t) = 0\) for all \(t\) if \(\lambda_B = 0\). Otherwise, although \(\delta\) is zero at the beginning, it can reach a non-zero value at the end of the decision-making process. Such revivals are likely to occur due to any interconversion between the quantum coherence and probabilities during the closed or open system dynamics of the mental state, described by the completely-positive trace-preserving map \(\varepsilon_t\).

\section{Information-Theoretical Concepts} \label{Sec::QInfo}

We show in Sec.~\ref{Sec::AnalResuls} that for the prediction under uncertainty to violate STP, it must involve quantum coherence. But what is the relationship between uncertainty and quantum coherence? We will discuss this in what follows from the perspective of quantum information theory.

\subsection{Quantum Uncertainty}

The amount of disorder, uncertainty, or unpredictability of a system prepared in the state \(\rho\) can be quantified by \textit{von Neumann entropy}~\cite{Neumann1927}, which is defined as
\begin{equation}\label{Eq::Entropy}
S[\rho] = - \mathrm{tr}\![\rho \, \log_2 \rho] .
\end{equation}

\(S[\rho]\) vanishes for pure states, that is, there is no uncertainty or randomness when \(\rho\) can be written as the self-outer product of a single state vector. On the other hand, \(S[\rho]\) is maximal for the mixed states whose eigenvalues comprise a uniform probability distribution. This corresponds to the highest degree of randomness in a quantum state.

Substitution of the eigendecomposition \(\rho = \sum_j \epsilon_j \ketbra{\epsilon_j}{\epsilon_j}\) into \eqref{Eq::Entropy} results in \(S[\rho] = - \sum_j \epsilon_j \log_2 \epsilon_j\),
where \(\{\epsilon_j, \ket{\epsilon_j}\}\) constitute the eigenspectrum of \(\rho\). Accordingly, \(S[\rho]\) can be regarded as the quantum generalization of Shannon information entropy~\cite{Shannon1, Shannon2}, which means that it is also related to the average amount of quantum information contained in \(\rho\)~\cite{1995_Schumacher}.

\subsection{Quantum Coherence}

Quantum coherence is a manifestation of quantum nonlocality and is quantified by the amount of quantum superposition possessed in a state with respect to a fixed orthonormal basis. Here, we will discuss two proper measures of quantum coherence, both proposed in Ref~\cite{Baum14}. The first one is the \textit{$l_1$ norm of coherence}, defined as
\begin{equation}\label{Eq::l1coh}
    C_{l_1}[\rho] = \sum_{i\neq j} |\rho_{ij}| ,
\end{equation}
while the second one is the \textit{relative entropy of coherence}
\begin{equation}
    C_{RE} [\rho] = \min_{\varsigma \in \mathrm{IC}} S[\rho || \varsigma]  = S(\rho_d) - S(\rho) ,
\end{equation}
where \(S[\rho || \varsigma]\) is the \textit{relative von Neumann entropy} of \(\rho\) with respect to \(\varsigma\), \(\mathrm{IC}\) is the set of all incoherent states, and \(\rho_d\) is the diagonal part of the density matrix.

\(C_{l_1}\) takes into account distinct off-diagonal terms in \(\rho\) independently of each other, and each off-diagonal term is associated with a nonlocality among two basis states. Conversely, \(C_{RE}\) does not discriminate between the off-diagonal terms and highlights the overall nonclassicality, showing a holistic picture. It also corresponds to how much the loss of all coherence would increase uncertainty in the system.

\subsection{Quantum Correlations}

Quantum correlations arise from quantum coherence shared across subsystems without being localized. Here, we will consider only a particular type, quantum entanglement.

The \textit{entropy of entanglement}~\cite{1996_EntOfEnt} which is defined as
\begin{equation}\label{Eq::EntOfEnt}
E_E[| \psi \rangle \langle \psi |] = (S \circ
\mathrm{tr}_{A(B)})[| \psi \rangle \langle \psi |] ,
\end{equation}
is the unique entanglement measure for pure bipartite states~\cite{1997_EntOfEnt}. Hence, the uncertainty in subsystems increases with increasing entanglement in the pure joint state.

Entanglement entropy can be extended to the mixed states by means of a convex roof. This extension is known as the \textit{entanglement of formation}~\cite{1996_EoF}
\begin{equation} \label{Eq::EoF1}
E_F[\rho] = \min \left( \sum_j r_j \, E_E[| \psi_j
\rangle \langle \psi_j |] \right) ,
\end{equation}
where the minimum is taken over all the possible pure state
decompositions of \(\rho\) that realize \(\rho = \sum_j
r_j \, | \psi_j \rangle \langle \psi_j | \).

\section{Numerical Results}\label{Sec::NumResuls}

Despite their connections to quantum coherence discussed in Sec.~\ref{Sec::QInfo}, it is not straightforward to show whether quantum uncertainty and entanglement also play a role in STP violation. In this section, we will investigate the dynamics of uncertainty and entanglement in the mental state of the player. To this aim, we will initialize the dynamics exemplarily by four initial states with different uncertainty and coherence distributions.

\subsection{Initial Mental States} \label{Sec::MentalStates}

We consider initially uncorrelated mental states
\begin{equation} \label{Eq::Rho:total}
\rho^{(\alpha)}(0) = \rho^{(\alpha)}_B(0) \otimes \rho^{(\alpha)}_A(0) ,
\end{equation}
where $\alpha = \{u, d, c\}$ as in Sec.~\ref{Sec::AnalResuls}. Each subsystem is described by only two parameters, $\lambda_{A/B}$ and $p_{A/B}$, as explained in Eqs.~(\ref{Eq::RhoA::Q}-\ref{Eq::RhoB::Q}) respectively. In the case of exact prediction at the beginning of the decision-making process, \(p_B\) is either zero or one and so, \(\mathrm{tr}_A[\rho^{(\alpha \neq u)}] = \ketbra{\alpha}{\alpha}_B\). Also, \(\rho^{(u)}_A(0) = \rho^{(d)}_A(0) = \rho^{(c)}_A(0)\) in order to make the initial action state independent of the uncertainty. We focus on four exemplary mental states as follows.

\textit{1) Classical-classical states} -- The first set of mental states that we discuss possess no quantum coherence initially localized either in prediction or action subsystem. However, there is maximum uncertainty in both subsystems. The resulting mental states labeled by Case \(1\) in Table~\ref{Table::InitialStates} correspond to the initial states previously used in a CPT Model by Pothos and Busemeyer~\cite{Buse09}. Here, we also study different amounts of uncertainty by decreasing the uniformity of the marginal probability distributions (see Case \(1^*\) in Table~\ref{Table::InitialStates}).

\textit{2) Classical-quantum states} -- Unlike the previous case, we inject quantum coherence initially into the action subsystem. As shown in Case \(2\) in Table~\ref{Table::InitialStates}, the injection creates maximum local coherence in this subsystem by removing all uncertainty.

\textit{3) Quantum-classical states} -- In this case, we introduce quantum coherence to the initial prediction state. By maximizing coherence injection, we can make the state \(\rho^{(u)}_B\) pure, completely eliminating uncertainty in the prediction. Consequently, this approach does not correspond to a lack of exact prediction at the beginning of the decision-making process. This makes Case \(3\) an improper mental state. However, if the injected quantum coherence is only half of the maximum amount, a significant degree of uncertainty (81\%) continues to persist in the prediction  (refer to Case \(3^*\) in Table~\ref{Table::InitialStates}).

\textit{4) Quantum-quantum states} -- The final set of mental states we are examining is characterized by the quantum coherence initially localized in both the prediction and action subsystems. In the case of a maximum amount of quantum coherence, there is no uncertainty present in either of these subsystems. This particular mental state is referred to as Case \(4\) in Table~\ref{Table::InitialStates}, representing the initial states previously employed in a QPT model developed by Pothos and Busemeyer~\cite{Buse09}. However, it is not considered a suitable mental state like Case \(3\). Therefore, we investigate states that exhibit a combination of coherence and uncertainty in the prediction by reducing the magnitude of \(\lambda_B\), denoted as Case \(4^*\) in Table~\ref{Table::InitialStates}.

\begin{table}[t]
\caption{The parameters, $p_{B/A}$ and $\lambda_{B/A}$, that characterizes different initial states under consideration. Quantum information content and uncertainty amount in the prediction and action are quantified by von Neumann entropy ($S$) and $l_1$ norm of coherence ($C_{l_1}$) for 7 different cases. Note that the cases with vanishing \(S(B)\) for \(\alpha = u\) cannot be considered a suitable mental state.}
\centering
\begin{tabular}{{|c|c|c|c|c|c|c|c|c|c|}}
\hline
Case No. & \(\alpha\) & \(p_B\) & \(\lambda_B\) & \(C_{l_1}(B)\) & \(S(B)\) & \(p_A\) & \(\lambda_A\) & \(C_{l_1}(A)\) & \(S(A)\) \\
\hline
 & \(u\) & $1/2$ & $0$ & 0 & 1 & $1/2$ & $0$ & 0 & 1 \\\cline{2-10}
1 & \(d\) & $1$ & $0$ & 0 & 0 & $1/2$ & $0$ & 0 & 1 \\\cline{2-10}
& \(c\) & $0$ & $0$ & 0 & 0 &$1/2$ & $0$ & 0 & 1 \\\hline

 & \(u\) & $1/3$ & $0$ & 0 & 0.92 & $3/5$ & $0$ & 0 & 0.97 \\\cline{2-10}
1* & \(d\) & $1$ & $0$ & 0 & 0 & $3/5$ & $0$ & 0 & 0.97 \\\cline{2-10}
& \(c\) & $0$ & $0$ & 0 & 0 &$3/5$ & $0$ & 0 & 0.97 \\\hline

& \(u\) & $1/2$ & $0$ & 0 & 1 & $1/2$ & $1/2$ & 1 & 0 \\\cline{2-10}
2 & \(d\) & $1$ & $0$ & 0 & 0 & $1/2$ & $1/2$ & 1 & 0 \\\cline{2-10}
& \(c\) & $0$ & $0$ & 0 & 0 & $1/2$ & $1/2$ & 1 & 0 \\\hline

& \(u\) & $1/2$ & $1/2$ & 1 & 0 & $1/2$ & $0$ & 0 & 1 \\\cline{2-10}
3 & \(d\) & $1$ & $0$ & 0 & 0 & $1/2$ & $0$ & 0 & 1 \\\cline{2-10}
& \(c\) & $0$ & $0$ & 0 & 0 & $1/2$ & $0$ & 0 & 1 \\\hline

& \(u\) & $1/2$ & $1/4$ & 0.5 & 0.81 & $1/2$ & $0$ & 0 & 1 \\\cline{2-10}
3* & \(d\) & $1$ & $0$ & 0 & 0 & $1/2$ & $0$ & 0 & 1 \\\cline{2-10}
& \(c\) & $0$ & $0$ & 0 & 0 & $1/2$ & $0$ & 0 & 1 \\\hline

& \(u\) & $1/2$ & $1/2$ & 1 & 0 & $1/2$ & $1/2$ & 1 & 0\\\cline{2-10}
4 & \(d\) & $1$ & $0$ & 0 & 0 & $1/2$ & $1/2$ & 1 & 0 \\\cline{2-10}
& \(c\) & $0$ & $0$ & 0 & 0 & $1/2$ & $1/2$ & 1 & 0 \\\hline

& \(u\) & $1/2$ & $i/4$ & 0.5 & 0.81 & $1/2$ & $1/2$ & 1 & 0\\\cline{2-10}
4* & \(d\) & $1$ & $0$ & 0 & 0 & $1/2$ & $1/2$ & 1 & 0 \\\cline{2-10}
& \(c\) & $0$ & $0$ & 0 & 0 & $1/2$ & $1/2$ & 1 & 0 \\\hline
\end{tabular}
\label{Table::InitialStates}
\end{table}

\subsection{Quantum Dynamics}

Here and in what follows, our primary focus revolves around the specific dynamics of mental states presented in Ref.~\cite{Buse09}. The purpose of our investigation is to explore the potential influence of quantum uncertainty and entanglement on the violation of STP. In particular, we characterize the time evolution of the mental state by a unitary dynamics
\begin{equation}\label{Eq::Unitary}
    \rho(t)= \boldsymbol{U}(t) \rho(0) \boldsymbol{U}^{\dagger}(t) \, ,
\end{equation}
which is generated by a Hamiltonian operator \(\boldsymbol{H}\) as below:
\begin{equation}
    \boldsymbol{U}(t) \equiv \exp(-i\boldsymbol{H}t/\hbar) \, .
\end{equation}

Above, $\hbar$ is assumed to be 1, and the Hamiltonian operator is divided into two terms
\begin{equation}\label{Eq::Ham}
    \boldsymbol{H} = \boldsymbol{H}_A + \boldsymbol{H}_B \, ,
\end{equation}
such that each term describes a different interaction between the prediction and action of the player. The first interaction gives the prediction control over the action, which is governed by
\begin{equation}
\boldsymbol{H}_A = \sum_{i=d,c} |i\rangle \langle i|_B \otimes \boldsymbol{H}_{Ai} \, ,
\end{equation}
where
\begin{equation}
\boldsymbol{H}_{Ai} = \frac{1}{\sqrt{1 + \mu_i^2}}
    \begin{pmatrix}
	\mu_i & 1 \\
	1 & -\mu_i
	\end{pmatrix} \, ,
\end{equation}
with $\mu_i$ is a constant related to the pay-offs of actions \(d\) and \(c\) in the presence of prediction \(i\). In other words, \(\boldsymbol{H}_A\) is involved in the Hamiltonian specifically to mix the joint probabilities \(p_{id} = \langle i_B \otimes d_A |\rho(t)|i_B \otimes d_A\rangle\) and \(p_{ic} = \langle i_B \otimes c_A |\rho(t)|i_B \otimes c_A\rangle\) as a function of \(\mu_i\) and \(t\). It also manipulates the quantum coherence carried in the mental state. However, it cannot generate any quantum entanglement between initially separable prediction and action states.

We assume that both of the constants $\mu_d$ and $\mu_c$ are equal to $\mu$ as in~\cite{Buse09}. That is to say, the pay-offs are taken independently of the prediction.

The second term in the Hamiltonian~(\ref{Eq::Ham}) reads
\begin{equation}
\boldsymbol{H}_B = \sum_{j=d,c} \boldsymbol{H}_{Bj} \otimes |j\rangle \langle j|_A \, ,
\end{equation}
where
\begin{equation}
\boldsymbol{H}_{Bj} = - \frac{\gamma}{\sqrt{2}}
    \begin{pmatrix}
	\nu_j & 1 \\
	1 & -\nu_j
	\end{pmatrix} \, ,
\end{equation}
with \(\nu_d = -\nu_c = 1\). \(\boldsymbol{H}_B\) mixes the joint probabilities \(p_{dj} = \langle d_B \otimes j_A |\rho(t)|d_B \otimes j_A\rangle\) and \(p_{cj} = \langle c_B \otimes j_A |\rho(t)|c_B \otimes j_A\rangle\) as a function of \(\gamma\) and \(t\). This reverses the direction of the control between the action and prediction of the player. Furthermore, unlike \(\boldsymbol{H}_A\), \(\boldsymbol{H}_B\) does not only adjust the quantum coherence carried in the mental state but also builds quantum entanglement between initially separable prediction and action states.

Summing up the interactions described by \(\boldsymbol{H}_A\) and \(\boldsymbol{H}_B\) gives us the total Hamiltonian, which gives rise to the unitary time evolution in Eq.~(\ref{Eq::Unitary}). However, when the player has made up her/his mind to choose one action, the final stage of the decision-making process is modeled by a sudden measurement operation. We imagined that a projective measurement would be performed on the action state. This measurement destroys all the quantum coherence and correlations in the action state and collapses it into the product state \(|j\rangle_B\) with a probability of \(p_{dj} + p_{cj}\).

\begin{table}[t]
\caption{The mean uncertainty in the mental states being examined are evaluated over the time interval from \(t=0\) to \(2 \pi\). The amount of quantum uncertainty is measured through von Neumann entropy \(S\). The STP  column indicates whether each case adheres to or violates the STP during the time interval of \((0, 2 \pi)\). The last column gives the total correlations shared between the action and prediction states, which is given by \(I(A:B) = S(A) + S(B) - S(AB)\)}
\centering
\begin{tabular}{{|c|c|c|c|c|c|c|c|c|}}
\hline
Case & STP & \(\alpha\)  & \(S(B)\) & \(S(A)\) & \(S(AB)\) & \(I(A:B)\) \\
\hline
 & & \(u\)  & 1 & 1 & 2 & 0   \\\cline{3-7}
1 & \cmark & \(d\)  & 0.65 & 1 & 1 & 0.65 \\\cline{3-7}
& & \(c\)  & 0.65 & 1 & 1 & 0.65 \\\hline

 & & \(u\)  & 0.96 & 0.99 & 1.89 & 0.06 \\\cline{3-7}
1* & \cmark & \(d\) & 0.66 & 0.99 & 0.97 & 0.68  \\\cline{3-7}
& & \(c\)  & 0.66 & 0.98 & 0.97 & 0.66 \\\hline

 & & \(u\) & 0.88 & 0.70 & 1 & 0.59 \\\cline{3-7}
2 & \cmark & \(d\) & 0.41 & 0.40 & 0 & 0.81  \\\cline{3-7}
& & \(c\)  & 0.50 & 0.50 & 0 & 1.00 \\\hline

 & & \(u\)  & 0.81 & 0.80 & 1 & 0.61  \\\cline{3-7}
3 & \xmark & \(d\)  & 0.67 & 1 & 1 & 0.67  \\\cline{3-7}
& & \(c\) & 0.67 & 1 & 1 & 0.67 \\\hline

 & & \(u\) & 0.96 & 0.95 & 1.81 & 0.10 \\\cline{3-7}
3* & \xmark & \(d\) & 0.67 & 1 & 1  & 0.67 \\\cline{3-7}
& & \(c\)  & 0.67 & 1 & 1 & 0.67 \\\hline

 & & \(u\)  & 0.53 & 0.53 & 0 & 1.05 \\\cline{3-7}
4 & \xmark & \(d\) & 0.41 & 0.41 & 0 & 0.81
 \\\cline{3-7}
& & \(c\)  & 0.50 & 0.50 & 0 & 1.00 \\\hline

 & & \(u\) & 0.76 & 0.61 & 0.81   & 0.56 \\\cline{3-7}
4* & \xmark & \(d\) & 0.41 & 0.41 & 0
 & 0.81 \\\cline{3-7}
& & \(c\) & 0.50 & 0.50 & 0 & 1.00  \\\hline
\end{tabular}
\label{Table::MeanQuantumUncertainty}
\end{table}

\begin{table}[t]
\caption{The mean quantum properties of the mental states being examined are evaluated over the time interval from \(t=0\) to \(2 \pi\). The amount of quantum coherence is assessed using two masures:  $l_1$ norm of coherence ($C_{l_1}$) and relative entropy of coherence ($C_{RE}$). Meanwhile, quantum entanglement is quantified by using entanglement of formation ($E_F$). The STP  column indicates whether each case adheres to or violates the STP during the time interval of \((0, 2 \pi)\).}
\centering
\begin{tabular}{{|c|c|c|c|c|c|c|c|c|c|}}
\hline
Case & STP & \(\alpha\)  & \(C_{l_1}(B)\) & \(C_{l_1}(A)\) & \(C_{l_1}(AB)\) & \(C_{RE}(AB)\) & \(E_{F}(AB)\) \\
\hline
 & & \(u\)  & 0 & 0 & 0 & 0 & 0  \\\cline{3-8}
1 & \cmark & \(d\)  & 0.38 & 0 & 1.17 & 0.83 & 0 \\\cline{3-8}
& & \(c\)  & 0.38 & 0 & 1.17 & 0.83 & 0 \\\hline

 & & \(u\)  & 0.15 & 0.10 & 0.52 & 0.09 & 0 \\\cline{3-8}
1* & \cmark & \(d\) & 0.39 & 0.09 & 1.30 & 0.84 & 0.04 \\\cline{3-8}
& & \(c\)  & 0.40 & 0.12 & 1.32 & 0.85 & 0.03 \\\hline

 & & \(u\) & 0.31 & 0.43 & 1.40 & 0.86 & 0.08 \\\cline{3-8}
2 & \cmark & \(d\) & 0.66 & 0.64 & 2.39 & 1.60 & 0.41 \\\cline{3-8}
& & \(c\)  & 0.58 & 0.59 & 2.34 & 1.57 & 0.50 \\\hline

 & & \(u\)  & 0.42 & 0.39 & 1.36 & 0.85 & 0.11 \\\cline{3-8}
3 & \xmark & \(d\)  & 0.38 & 0 & 1.17 & 0.83 & 0 \\\cline{3-8}
& & \(c\) & 0.38 & 0 & 1.17 & 0.83 & 0 \\\hline

 & & \(u\) & 0.21 & 0.19 & 0.68 & 0.15 & 0\\\cline{3-8}
3* & \xmark & \(d\) & 0.38 & 0 & 1.17 & 0.83 & 0 \\\cline{3-8}
& & \(c\)  & 0.38 & 0 & 1.17 & 0.83 & 0\\\hline

 & & \(u\)  & 0.55 & 0.64 & 2.59 & 1.72 & 0.53 \\\cline{3-8}
4 & \xmark & \(d\)  & 0.66 & 0.64 & 2.39 & 1.59 & 0.41
 \\\cline{3-8}
& & \(c\)  & 0.58 & 0.59 & 2.34 & 1.57 & 0.50 \\\hline

 & & \(u\) & 0.48 & 0.54 & 1.72 & 1.00 & 0.14  \\\cline{3-8}
4* & \xmark & \(d\) & 0.66 & 0.64 & 2.39 & 1.59 & 0.41
 \\\cline{3-8}
& & \(c\) & 0.58 & 0.59 & 2.34 & 1.57 & 0.50  \\\hline
\end{tabular}
\label{Table::MeanQuantumInfo}
\end{table}

\subsection{Uncertainty, Coherence, and Correlations}

\begin{figure}[t]
    \centering
    \includegraphics[width=9cm]{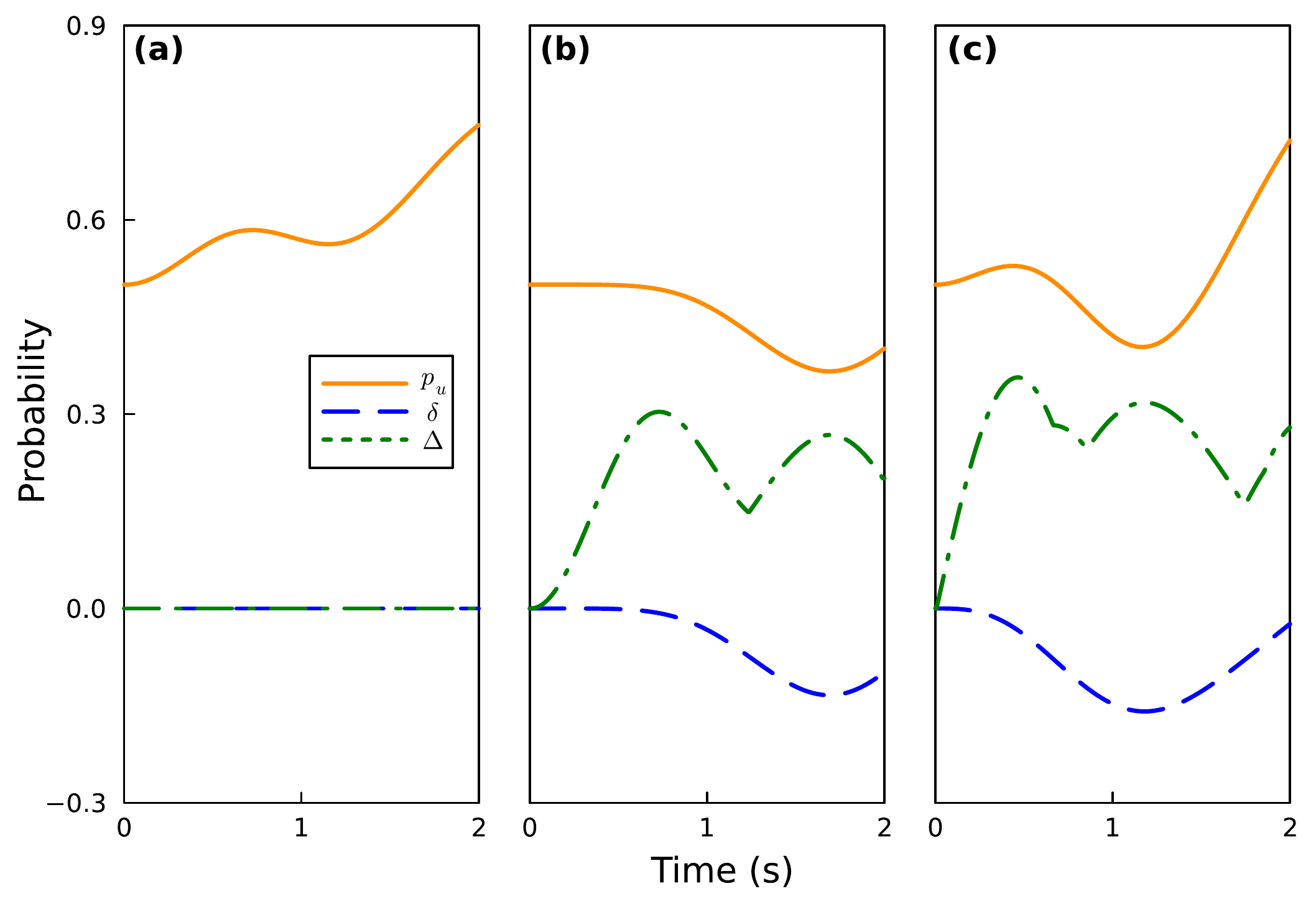}
    \caption{Time-dependent behavior of the probabilistic quantities \(p_u \equiv p_{A|B=u}\), \(\delta\), and \(\Delta \equiv \sum_{i} |\chi_{ii}|\) are shown by solid orange, dashed blue, and dotted green curves for (a) Case 2, (b) Case 3\(^*\), and (c) Case 4\(^*\)}.
    \label{Fig::Prob}
\end{figure}

In this section, we will analyze the unitary evolutions of the initial mental states presented in Table~\ref{Table::InitialStates}, considering the Hamiltonian defined in Eq.~\eqref{Eq::Ham}. Our previous investigation in Sec.~\ref{Sec::AnalResuls} established that in order to observe a violation of STP, it is necessary for \(\delta\) to have a non-zero value. This condition implies the presence of quantum coherence in the initial prediction, and it is an analytical result that remains unaffected by the dynamics of the mental state. To further validate this finding, we have conducted numerical simulations taking \(\gamma = 1.74\) and \(\mu_d = \mu_c = 0.59\), the results of which are summarized in Tables~\ref{Table::MeanQuantumUncertainty}~and~\ref{Table::MeanQuantumInfo}, as well as depicted in Fig.~\ref{Fig::Prob}. Consistently with our analytical analysis, we observe that STP is only violated when the initial prediction exhibits quantum coherence (Cases \(3\), \(3^*\), \(4\), and \(4^*\)).

Pothos and Busemeyer attempted to elucidate the violation of STP in Case 4 by interpreting the notion of quantum coherence as a kind of prediction uncertainty~\cite{Buse09}. However, as explained in Sec.~\ref{Sec::MentalStates} and illustrated in Table~\ref{Table::InitialStates}, Case 4 cannot be considered an appropriate mental state because of the absence of uncertainty in the initial prediction. Could it be possible that the temporal progression of uncertainty, rather than the initial level of uncertainty itself, plays a role in STP violation? This inquiry arises from the conceptual relationship between coherence and uncertainty discussed in Sec.~\ref{Sec::QInfo}. To investigate this possibility, we calculated the dynamics of local and global uncertainties in the mental states given in Table~\ref{Table::InitialStates}. The average values of these uncertainties are summarized in Table~\ref{Table::MeanQuantumUncertainty}. Importantly, it seems impractical to differentiate between cases that violate STP and those that do not solely based on uncertainty dynamics.

The discussion in Sec.~\ref{Sec::QInfo} raises one more question regarding the violation of STP. Could the closed system dynamics generate entanglement between the prediction and action, potentially impacting the violation? However, a thorough examination of the findings summarized in Table~\ref{Table::MeanQuantumInfo} reveals that quantum entanglement does not play a role in this violation. In Case \(1\), where STP remains intact, entanglement does not arise. Conversely, in Case \(1^*\) and Case \(2\), entanglement between prediction and action can occur. A similar pattern emerges in the scenarios where STP is violated. For instance, in Case \(3^*\), the mental state remains separable, while entanglement is observed in the remaining cases. Overall, these results indicate that while the local quantum coherence possessed by the initial prediction can be converted to quantum entanglement between prediction and action during the time evolution, it does not directly contribute to the violation.

When comparing the results from Table~\ref{Table::InitialStates} and Table~\ref{Table::MeanQuantumInfo}, another important point emerges in terms of the dynamics of the mental state governed by Eq.~\eqref{Eq::Ham}. It is capable of converting the initial quantum coherence in both prediction and action into quantum entanglement. Furthermore, what is even more interesting is that even if there is no quantum coherence in the initial mental state, the action and prediction can still become entangled. Case \(1^*\) exemplifies this phenomenon, where the initial action and prediction states are described by classical, yet nonuniform probability distributions. Remarkably, if the dynamics of the mental state are quantum in nature, quantum coherence and entanglement can still emerge.

In conclusion, it can be asserted that prediction uncertainty or action-prediction entanglement do not contribute to the violation of STP. The quantum coherence present in the initial prediction, as discussed in Sec.~\ref{Sec::AnalResuls}, undergoes a conversion into probabilities during the evolution of the mental state. This mechanism elucidates the observed violation of STP in the QPT model originally proposed in Ref.~\cite{Buse09}. Furthermore, we substantiate our analytical discussion in Sec.~\ref{Sec::AnalResuls} with numerical calculations presented in Fig.~\ref{Fig::Prob}. The diagonal elements of the matrix \(\chi\) in Eq.~\eqref{Eq::Ksi0}, as depicted, exhibit non-zero values over time in cases where STP is violated. Notably, a significant finding emerges when examining the violation of STP: it initiates at the point where a change occurs in the behavior of \(\Delta\), the summed magnitudes of these probabilistic quantities. Although beyond the scope of this paper, we intend to investigate this intriguing result in future studies.

\section{Outlook}\label{Sec::Outlook}

In this paper, we provide an integrated perspective on the classical and quantum decision-making models proposed by Busemeyer and Pothos in Ref.~\cite{Buse09}. Our approach leverages the unified framework of GPTs, enabling us to represent both classical and quantum mental states as density matrices. To capture the temporal dynamics of these mental states, we adopt the Hamiltonian utilized in the QPT model. By examining the data presented in Table~\ref{Table::InitialStates} and Table~\ref{Table::MeanQuantumInfo}, we observe that this Hamiltonian has the capability to transform a classical initial mental state into a quantum one. To address this issue, we aim to develop an open system approach that preserves the classical nature of all initial states. This can be achieved by establishing a master equation that incorporates the Hamiltonian presented in Eq.~\eqref{Eq::Ham}.

Busemeyer and Pothos explained the observed violation of the STP in the QPT model through the second-order interference effect~\cite{Buse09}. However, they arrived at this conclusion by representing the absence of exact prediction with an inappropriate mental state. In this study, we examined the dynamics of various mental systems. Both theoretically and numerically, we demonstrated that the source of the interference leading to STP violation is the quantum coherence in the initial prediction. We also uncovered the underlying mechanism: the quantum coherence in the mental state of the initial prediction undergoes a transformation into probabilities during temporal evolution. Furthermore, we showed that the quantum uncertainty emerging in the prediction over time or the quantum entanglement established between prediction and action states does not play a role in this mechanism.

The initial mental state utilized in the QPT model of Busemeyer and Pothos~\cite{Buse09} was designated as Case \(4\) in Table~\ref{Table::InitialStates}. In addition to its lack of prediction uncertainty, as illustrated in Table~\ref{Table::MeanQuantumInfo}, this state enables quantum entanglement to manifest between the action and prediction subsystems over time. However, the question of whether it is feasible to partition a mental system during the decision-making process in a manner that attributes nonlocality remains unanswered. To avoid venturing into speculative discussions, our investigation proposes replacing the mental state in the QPT model with a state, such as the one in Case \(3^*\), that consistently remains separable.

The GPT framework we adopt here offers us a wealth of insights beyond analyzing current probabilistic models of decision-making processes. As previously mentioned in the Introduction, phenomena such as second-order interference are not limited to QPT alone. There are many non-quantum or beyond-quantum GPTs where similar phenomena can be observed. In fact, as initially proposed by Sorkin in Ref.~\cite{Sork94}, there could be GPTs capable of accommodating interference phenomena of three or higher orders. Considering this, could these types of GPTs outperform QPT in effectively modeling human decision-making mechanisms in games like PDG?

We can elucidate the concept of higher-order interference by referring to the famous double-slit experiment as in Ref.~\cite{Muller21}. Suppose we have two slits and consider the probability of detecting a particle on the screen when both slits are open, denoted $P_{12}$. Similarly, we have probabilities $P_1$ and $P_2$ for detecting a particle when only slit 1 or slit 2 is open, respectively. According to CPT, we expect the equality $P_{12} = P_1 + P_2$, whereas QPT predicts an inequality $P_{12} \neq P_1 + P_2$. Now, let's expand this experiment by introducing a third slit. Interestingly, both QPT and CPT yield the same equation: $P_{123} = P_{12} + P_{13} + P_{23} - P_1 - P_2 - P_3$. However, when we consider a GPT that accounts for third-order interference, we find that $P_{123} \neq P_{12} + P_{13} + P_{23} - P_1 - P_2 - P_3$. An analogy can be drawn between the choices of players in the PDG and the slits in the double-slit experiment. Here, we contemplate what would occur if we extend the PDG to include three choices for each player. Would we observe an equality that aligns with both CPT and QPT? Or would we witness a violation that cannot be explained by either CPT or QPT?

If experiments conducted with a three-choice PDG reveal violations that cannot be captured by the QPT model, it becomes imperative to move on from the density matrix formalism to the density cube formalism~\cite{Daki14}. Our objective in this context is to extend the QPT model, initially proposed by Busemer and Pothos and examined here through quantum information theoretical methods, to encompass the three-choice PDG. By achieving this generalization, we can gain valuable insights into whether additional frameworks beyond QPT are necessary to effectively model decision-making processes.

\nocite{*}


%

\end{document}